# Numerical study on the effects of fluid properties in electrohydrodynamic pulsating jet under constant voltage

Yaohui Lu[1], Songyi Zhong[1,2,a)], Kai Leong Chong[3], Yang Yang[1], Tao Yue[1], Quan Zhang[1,2], and Long Li[1]

[1]School of Mechatronic Engineering and Automation, Shanghai University, Shanghai 200444, China

[2]School of Artificial Intellgence, Shanghai University, Shanghai 200444, China

[3]Shanghai Institute of Applied Mathematics and Mechanics, School of Mechanics and Engineering Science, and Shanghai Key Laboratory of Mechanics in Energy Engineering, Shanghai University, Shanghai 200072, China

[a)]Author to whom correspondence should be addressed: zhongsongyi@shu.edu.cn

**Abstract:** Pulsating jet is one of the common working modes in electrohydrodynamic printing (EHDP) that process is highly affected by operating parameters and material properties. In this paper, the processes of pulsating jet for liquids with different physical properties were investigated using numerical simulation. An electrohydrodynamic solver was established, and a charge flux restricting step was adopted to ensure a realistic distribution of free charges. Three various ejection regimes were observed in our simulations: oscillating cone (OC), choked jet (CJ), and stable cone-jet (SJ). We found that three dimensionless numbers relating to liquid properties determined the ejection regime: the Ohnesorge number, $Q_0 \varepsilon_r/Q$, and $Q_0/(QRe)$. Based on those dimensionless numbers, the roles of liquid properties on pulsating jet (OC and CJ) were analyzed. In OC, the break of the jet is due to the significant oscillation of the Taylor cone, which is mainly affected by viscosity and conductivity. In CJ, the jet emission is terminated by the excessive resistant force in the cone-jet transition region. For liquids with low and medium viscosity, the dominant resistant force is the polarization force or viscous force when $\varepsilon_r Re$ is larger or smaller than 1, respectively. For high viscosity liquids, the viscous force always becomes the major resistance. These results further reveal the physical mechanism of pulsating jet and can be used to guide its application.

# I. Introduction

The deformation and ejection of liquid drops under an electric field have been of great concern to researchers, and a variety of techniques including electrohydrodynamic printing[1] (EHDP) have been developed based on such phenomena. EHDP has many advantages including high resolution and a wide range of applicable materials[2], hence it has been widely used in the manufacture of various devices in recent years, such as flexible sensors[3–5], scaffolds[6,7], dielectric actuators[8], and microfluidic chips[9].

The device used for EHDP typically consists of a nozzle that is applied a high voltage and a grounded substrate. An electric field is generated between the nozzle and substrate, which drives the liquid drop hanging in the nozzle orifice to eject via several modes[10,11]. The stable cone-jet[12] and pulsating jet[13] are the two most commonly used ejection modes in EHDP. In stable cone-jet, a continuous jet is emitted from the drop apex and deposited on substrate to form linear structures. In pulsating jet, the liquid meniscus periodically ejects jets or drops that form dot structures on substrate. By adopting the pulsed voltage, the ejection frequency and pulsation volume of pulsating jet can be adjusted independently, thus achieving on-demand EHDP[14]. However, the physical mechanism of pulsating jet is complex, and the ejection process is heavily influenced by a lot of factors, including liquid properties and operating parameters. Many researchers have conducted experimental studies on pulsating jet, especially the impacts of operating parameters such as voltage and flow rate. Juraschek and Rollgen[13] investigated the pulsating jet process under various voltages and observed that the ejection frequency changed from unstable to stable as the voltage increased. Marginean et al.[15] observed in experiments that the pulsating frequency of meniscus was consistent with the oscillation frequency of charged droplet under high voltage. Chen et al.[16] and Choi et al.[17] both studied the pulsating jet process experimentally and found that the ejection frequency increased with voltage. Guo et al.[18] observed in their experiments that the liquid volume ejected in a single pulsation declined with increased voltage. Bober and Chen[19] derived scaling laws between the input flow rate and ejection frequency for two different pulsating regimes theoretically and verified them by experiments. Hijano et al.[20] uncovered a scaling law between the diameter of the ejected droplets and the input flow rate through experiments.

Besides operating parameters, the liquid properties also play vital roles in the pulsating jet process. Different liquids may exhibit distinct ejection behaviors[19,21,22] under the same operating parameters due to their various physical properties. Several researchers have conducted experiments to study the effects of physical properties[18,23] on pulsating jet. However, a systematical investigation on this issue is inconvenient to perform using experiments due to the individual adjustment of each property is difficult. With the development of computational technology, numerical simulation has become an important method for overcoming the limitation of experiments and is widely applied in electrohydrodynamic studies, including spray[24–26], printing[27,28], and droplet manipulation[29–31]. More recently, researchers have carried out numerical studies on pulsating jet based on the Taylor-Melcher leaky dielectric model[32,33]. Guan's research team established two-dimensional axis-symmetric numerical models and investigated pulsating jet under constant voltage[34,35] and pulsed voltage[36]. They found several scaling relationships about the cone length, cone angle, and maximum swirling strength inside the cone. Cândido and Páscoa[37] developed a three-dimensional model for the EHD jet flow and analyzed the whipping dynamics of the jet. They speculated that the radial instability of jet is associated with the charge transport on the surface. According to the existing references, it is worth noting that most numerical studies on pulsating jet have not deeply discussed the role of liquid properties.

In this work, we investigate the role of liquid properties on pulsating jet under constant voltage by numerical simulation. A solver is developed based on the open-source software OpenFOAM. We have integrated a charge flux restricting step in the solver, which prevents free charges move across the liquid surface, thus ensuring more realistic solutions. The ejection processes of liquids with varying viscosities, permittivities, and conductivities are simulated. Three ejection regimes are identified, and the interface deformations and internal flow behaviors of each regime are discussed. The impacts of liquid properties are analyzed from the perspectives of cone oscillation and dominant resistant force in the cone-jet transition region.

## II. Numerical model

### A. Physical model

As shown in Fig. 1, we consider the most common configuration in practical applications which includes a nozzle and a substrate. The nozzle is conductive and applies a high voltage, while the substrate is grounded. The density, viscosity, relative permittivity, conductivity, and surface tension of liquid are denoted by $\rho$, $\mu$, $\varepsilon_r$, $K$, and $\sigma$, respectively. The density, viscosity, relative permittivity, and conductivity of the surrounding medium (air) are 1.29kg/m$^3$, 1.79×10$^{-5}$Pa*s, 1, and 10$^{-15}$S/m, respectively. The input flow rate is denoted by $Q$. The inner diameter ($D_i$) of the nozzle is 160μm, and the outer diameter ($D_o$) is 260μm. The length of the nozzle ($Z_0$) in the computational domain is 0.2mm. The distance between the nozzle orifice and the substrate ($H$) is 1.5mm. An axial symmetry coordinate with the origin located at the center of the nozzle is adopted. The radius of the liquid surface ($R_0$) is 80μm at the initial moment.

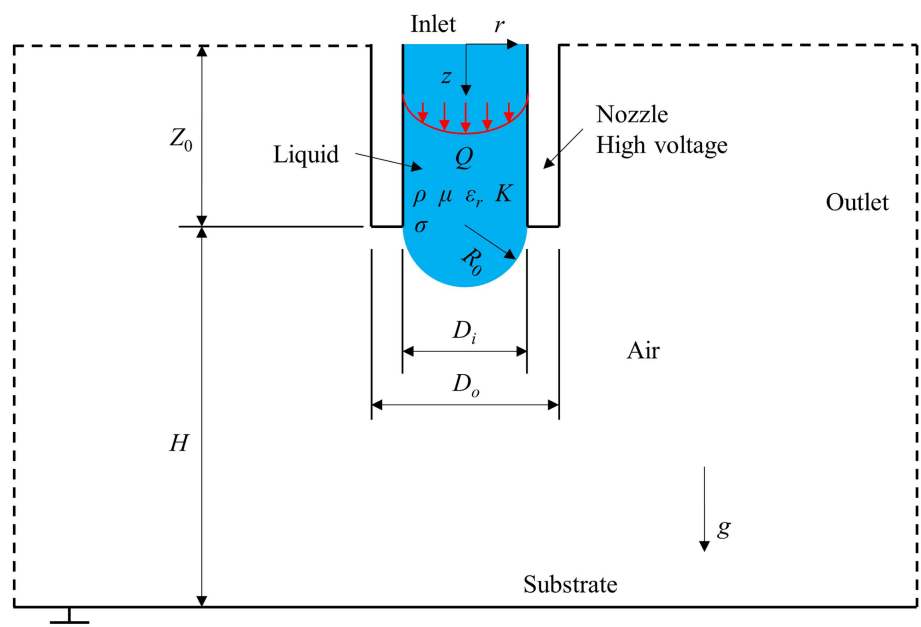

**Fig. 1** Sketch of the simulation domain.

We focus on three key dimensionless numbers related to liquid properties: relative permittivity ($\varepsilon_r$), dimensionless charge relaxation time ($\alpha$), and the Ohnesorge number ($Oh$). In addition, the density and surface tension of the liquid are fixed at 1000kg/m$^3$ and 0.02N/m, respectively. Thus, the dimensionless charge relaxation time is determined by the conductivity and the relative permittivity of the liquid:

$$\alpha = \frac{K\sqrt{\rho D^3/\sigma}}{\varepsilon_0 \varepsilon_r} \tag{1}$$

Where $D$ is the diameter of the droplet and is equal to the inner diameter of the nozzle, and $\varepsilon_0$ is the permittivity of vacuum. The liquid viscosity dictates the value of Ohnesorge number:

$$Oh = \frac{\mu}{\sqrt{\rho D \sigma}} \quad (2)$$

We have limited the range of liquid properties under investigation in our research. Specifically, the Ohnesorge number of the liquids does not exceed 1, which is a common viscosity range for drop-on-demand printing processes[38]. The relative permittivity is restricted to 2-180, while the dimensionless charge relaxation time is assumed larger than 1.

The same operation parameters will be applied in all simulations. The input flow rate ($Q$) is fixed at 20nl/s. The magnitude of the voltage can be evaluated using the electrical capillary number $Ca_E$:

$$Ca_E = \frac{\varepsilon_0 V^2 D_i}{\sigma (D_i \ln(4H/D_i))^2} \quad (3)$$

Where $V$ is the voltage applied on the nozzle. We only focus on the pulsating jet process under low voltages and thus the voltage is fixed at 1200V ($Ca_E \approx 0.3$).

**B. Numerical methods and the charge flux restriction step**

A numerical solver is established based on the open-source software OpenFOAM. The finite volume method (FVM) is adopted to solve the partial differential equations. The governing equations are based on the Taylor-Melcher leaky dielectric model and are given in the supplementary material due to the abundance of related references[30,31,39]. Here, we only introduce the special treatments for solving the charge conservation equation.

The space charge density $\rho_e$ can be calculated using the following equation[32,40]:

$$\frac{\partial \rho_e}{\partial t} + \nabla \cdot (\boldsymbol{u}\rho_e) = -\nabla \cdot (K\boldsymbol{E}) \quad (4)$$

Where $\boldsymbol{u}$, $K$, and $\boldsymbol{E}$ are the velocity vector, the conductivity of fluid, and the electric field vector, respectively. In the real world, free charges are limited to the liquid interface and cannot move across the interface unless an electrical breakdown occurs. However, solving Eq. (4) directly often leads to free charges crossing the interface and moving into the air phase due to numerical diffusion. Therefore, a charge flux restricting step is integrated to prevent the occurrence of such an unrealistic phenomenon. After discretization using the finite volume method, Eq. (4) is transformed into the following linear equation:

$$\Delta x \frac{\rho_e^t - \rho_e^{t-\Delta t}}{\Delta t} + \sum_f U_f \rho_e^t = -\sum_f K_f E_f^n \quad (5)$$

Where $\Delta x$, $\Delta t$, $\rho_e^t$, and $\rho_e^{t-\Delta t}$ represent the cell size, time step, and free charge density at the cell center for the current and last time step, respectively. $U_f$, $K_f$, and $E_f^n$ are the normal velocity, conductivity, and normal electric field at the center of a cell face, respectively. The flux of electric charge on the cell face consists of two components: the current induced by convection ($U_f \rho_e^t$), and the current induced by conduction ($K_f E_f^n$). The flux restriction only applies to the former component. For a cell (called the owner cell) in the computation domain, whether the flux of electric charge should be restricted depends on the liquid volume fraction at the cell center ($\varphi_o$) and its neighbor cell center ($\varphi_n$), as well as the normal velocity at the center of the shared cell face ($U_f$). The specific procedures of the charge flux restricting step are as follows:

(1) Obtain the liquid volume fraction at the center of the shared cell face ($\varphi_f$) through linear interpolation;

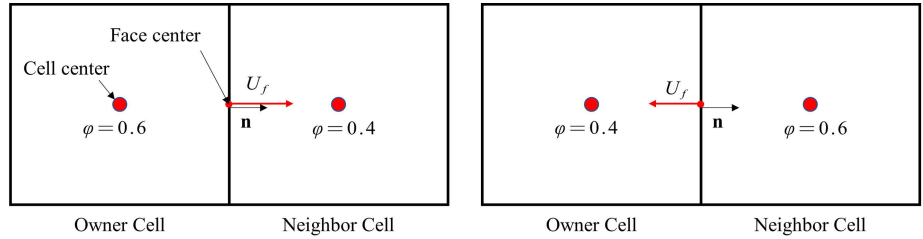

**Fig. 2** Two situations in which the convection flux of electric charge should be restricted.

(2) As shown in Fig. 2, if $\varphi_o > \varphi_n$ and $U_f > 0$, or $\varphi_o < \varphi_n$ and $U_f < 0$, the charges tend to move from the cell with a higher $\varphi$ to the cell with a lower $\varphi$. In this case, the normal velocity at the cell face center is multiplied by a coefficient $c_f$ and the resulting value is stored as a new normal velocity $U_f'$:

$$U_f' = c_f U_f \tag{6}$$

The value of the coefficient $c_f$ is related to $\varphi_f$ to ensure a smooth variation on the interface layer:

$$c_f = \begin{cases} 1, & \varphi_f \geq 0.5 \\ 2\varphi_f, & 0 < \varphi_f < 0.5 \\ 0, & \varphi_f = 0 \end{cases} \tag{7}$$

(3) The new normal velocity ($U_f'$) is used for solving Eq. (4), while the original normal velocity ($U_f$) is utilized for solving other equations. It is necessary to update $U_f'$ in each iteration and store it at the centers of cell faces.

The solver is developed by modifying the interFOAM solver, which is a built-in solver in

OpenFOAM. The implicit Euler scheme is used for time stepping. The gauss linear scheme is utilized for gradient interpolation, and the gauss linear upwind scheme is adopted for calculating convection terms. The classical PIMPLE algorithm is employed to solve the pressure-velocity coupling. Fig. S1 in the supplementary material summarizes the procedures involved in solving the equations.

## C. Model validation

We simulated the pulsating jet process based on the numerical model depicted in Fig. 1 and compared the numerical results with the experimental results by Guo et al.[18] to validate the accuracy of the numerical solver. The experimental setup and methodology are provided in detail in reference[18] and will not be described here. The inner and outer diameters of the nozzle used in experiments are 160μm and 260μm, respectively. The distance between the nozzle and the substrate is 1.5mm. The density of the working liquid is 1208.4kg/m$^3$, the viscosity is 60mPa*s, the conductivity is 6×10$^{-4}$S/m, the relative permittivity is 55.6, and the surface tension is 0.0645N/m. The input flow rate is 16.1nl/s, and the voltage is 2180V. The mesh used in the simulation is shown in Fig. S3 in the supplementary material, with a minimum mesh size of 1μm. We performed a mesh independence study (see supplementary material) and the results indicated that the solution had no significant change when the mesh size was reduced to below 1.6μm.

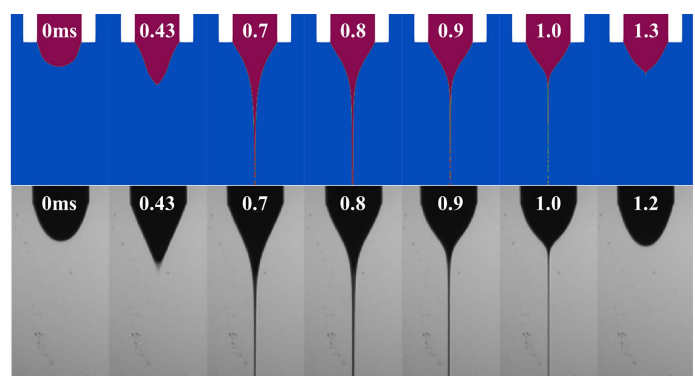

**Fig. 3** The comparison of the simulation results and the experimental images[18].

The comparison between numerical results and experimental images is shown in Fig. 3. After the voltage was applied at the initial moment, the meniscus started deforming and formed a Taylor cone[12] at approximately 0.43ms. Subsequently, a fine jet was ejected from the tip of the cone. The diameter of the jet gradually decreased until the jet eventually vanished. The similarity in the deformations of the liquid surfaces between the numerical results and the experiment indicates that the solver is capable of accurately simulating the pulsating jet process.

We also performed separate simulations for both opened and closed the flux restricting step to assess the effects of the proposed charge flux restricting step. As shown in Fig. 4(a) and (b), a few charges moved across the liquid interface and were carried by the airflow when the flux restricting step was closed. In contrast, the charges were entirely confined within the liquid interface when the flux restricting step was opened. This is further confirmed evidently in the distribution of free charges along the axis of symmetry in Fig. 4(c). Therefore, the charge flux restricting step successfully limited the charges to the liquid interface.

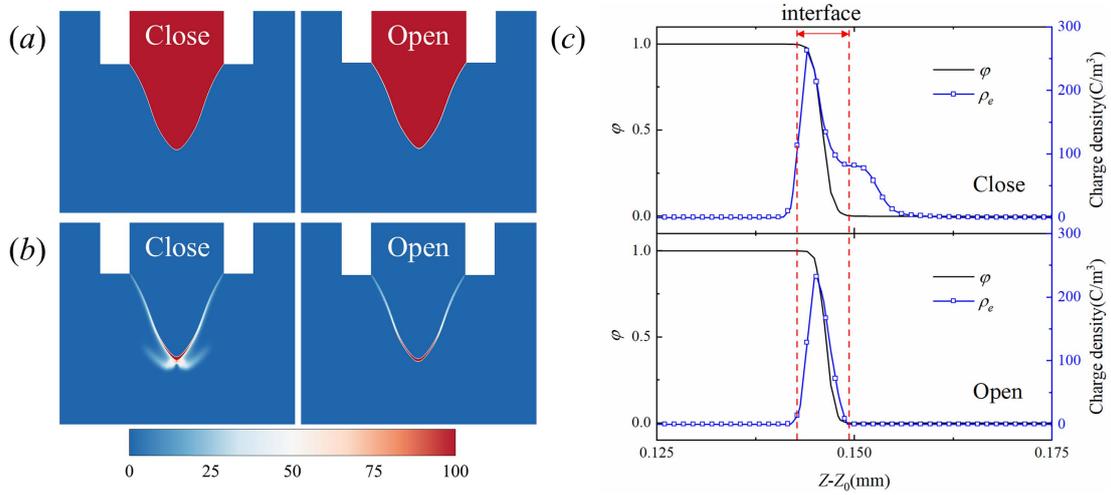

**Fig. 4** Simulation results with charge flux restriction closed and opened. (a) The liquid interface; (b) the charge distribution; and (c) the liquid volume fraction and charge density along the axis of symmetry.

## III. Results and discussion

The physical properties of liquids in our simulations are listed in Table. S1-S3 of the supplementary material. Three ejection regimes are observed in the results. The ejection process, interface deformation, and internal flow behavior of these regimes will be elucidated first. Subsequently, the parameters affecting the ejection regime are discussed. Finally, the roles of liquid properties during the jetting process are analyzed.

### A. Different ejection regimes

According to Bober and Chen's research[19], the pulsating jet can be divided into two regimes based on the relationship between the input flow rate and the minimum flow rate of liquid[41–43] (the minimum input flow rate required to form a stable cone-jet). As shown in Fig. 5, the ejection regime should be the oscillating cone with a high ejection frequency when the input flow rate exceeds the minimum flow rate. Conversely, the ejection regime should be the choked jet with a low ejection frequency when the input flow rate is below the minimum flow rate. In our simulations, we observed the oscillating cone, choked jet, and stable cone-jet regimes.

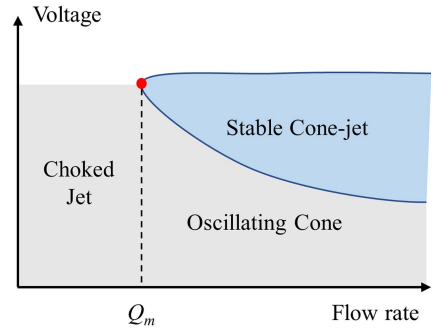

**Fig. 5** Ejection regimes distribution in the voltage-flow rate operation diagram.

1. Oscillating cone regime (OC)

The ejection process of the liquid with $Oh=0.1$, $\alpha=100$, and $\varepsilon_r=20$ is shown in Fig. 6(a). Under the action of electric forces, the liquid surface began to deform from a hemisphere (0ms) to a cone shape (0.19ms). A fine jet was emitted from the cone tip where the electric field was significantly strong (0.3ms). After a certain duration, the diameter of the cone-jet transition region continuously decreased until the jetting was suspended (0.53ms). Finally, the detached jet formed several liquid droplets, while the liquid cone gradually reverted to a hemisphere and prepared for the next ejection (0.57ms). As shown in Fig. 6(b), the deformation of the cone surface is significant in a certain period before the end of jetting, and the contact angle $\theta$ at the nozzle orifice noticeably increases as the cone surface rebounds. This ejection regime is considered as the oscillating cone regime (OC).

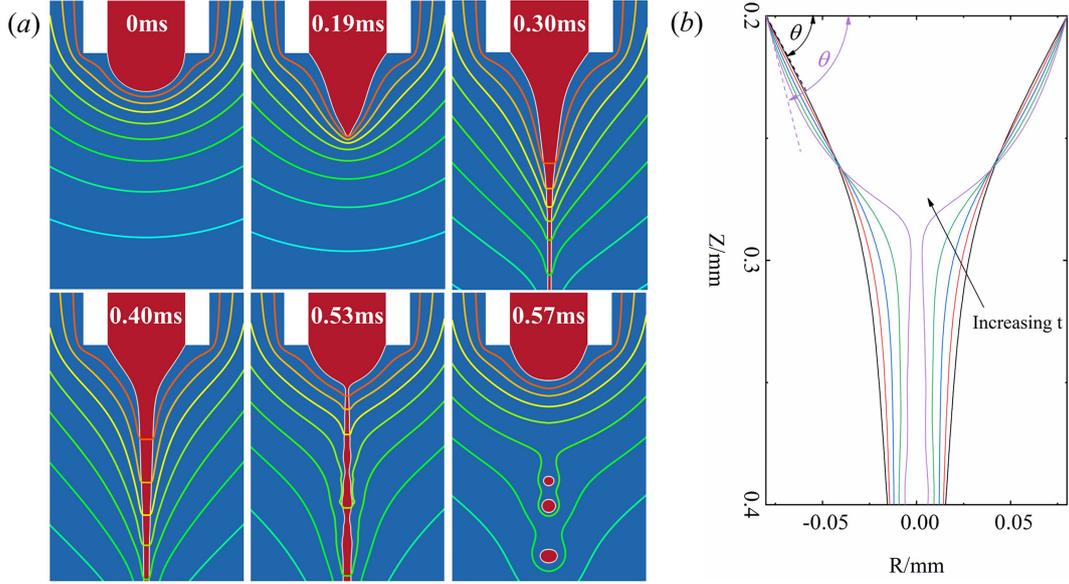

**Fig. 6** (a) The ejection process of the oscillating cone regime, the isolines represent the voltage; (b) the interface deformation of the oscillating cone regime during 0.33ms-0.53ms, the time interval between adjacent interfaces is 0.05ms, $\theta$ represents the contact angle of the liquid cone at the nozzle orifice.

2. Choked jet regime (CJ)

The ejection process of the liquid with $Oh=0.1$, $\alpha=10$, and $\varepsilon_r=20$ is shown in Fig. 7(a). Similar with the oscillating cone regime, a Taylor cone was formed at 0.23ms and a jet was ejected from its apex (0.4ms). The jet diameter decreased continuously until the jet vanished at 0.88ms. It can be observed from Fig. 7(b) that the entire cone has almost no oscillation during ejection, and the contact angle $\theta$ at the nozzle orifice remains constant. Due to the input flow rate being lower than the downstream emitted flow rate (equal to the minimal flow rate[19,41]), the diameter of the cone-jet transition region was decreased continuously as time evolved. The cone-jet transition region becomes unstable when its diameter is below a certain value[41] and then the jetting terminates. This ejection regime is considered as the choked jet regime (CJ), which has a longer ejection duration and lower frequency compared to the oscillating cone regime.

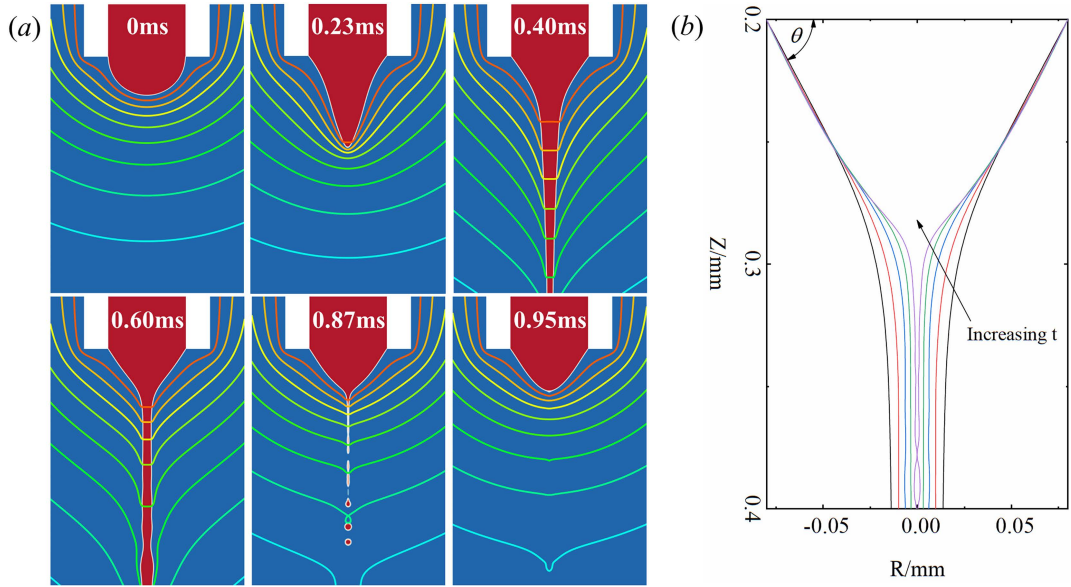

**Fig. 7** (a) The ejection process of the choked jet regime, the isolines represent the voltage; (b) the interface deformation of the choked jet regime during 0.47ms-0.87ms, the time interval between adjacent interfaces is 0.1ms, $\theta$ represents the contact angle of the liquid cone at the nozzle orifice.

3. Stable cone-jet regime (SJ)

The ejection process of the liquid with $Oh = 0.5$, $\alpha = 100$, and $\varepsilon_r = 40$ is depicted in Fig. 8. The formations of the Taylor cone and jet are as same as the preceding regimes. After approximately 2ms, the variables in the computational domain almost no longer changed, indicating that the jet had achieved a stable state. This ejection regime is the stable cone-jet regime (SJ). The mechanisms of this regime have been well understood through extensive investigations in previous literatures and will not be discussed deeply in this work.

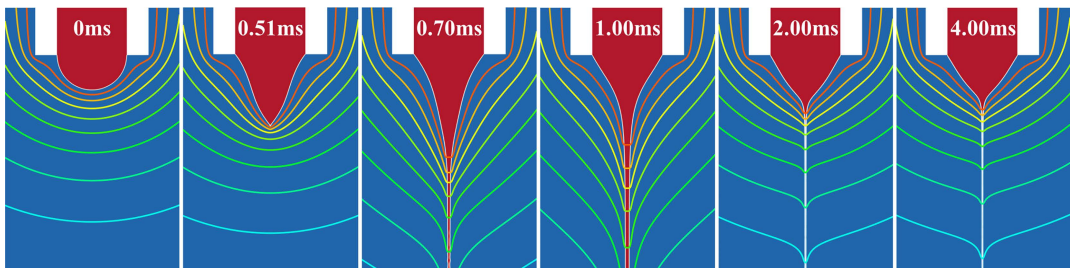

**Fig. 8** The ejection process of the stable cone-jet regime, the isolines represent the voltage.

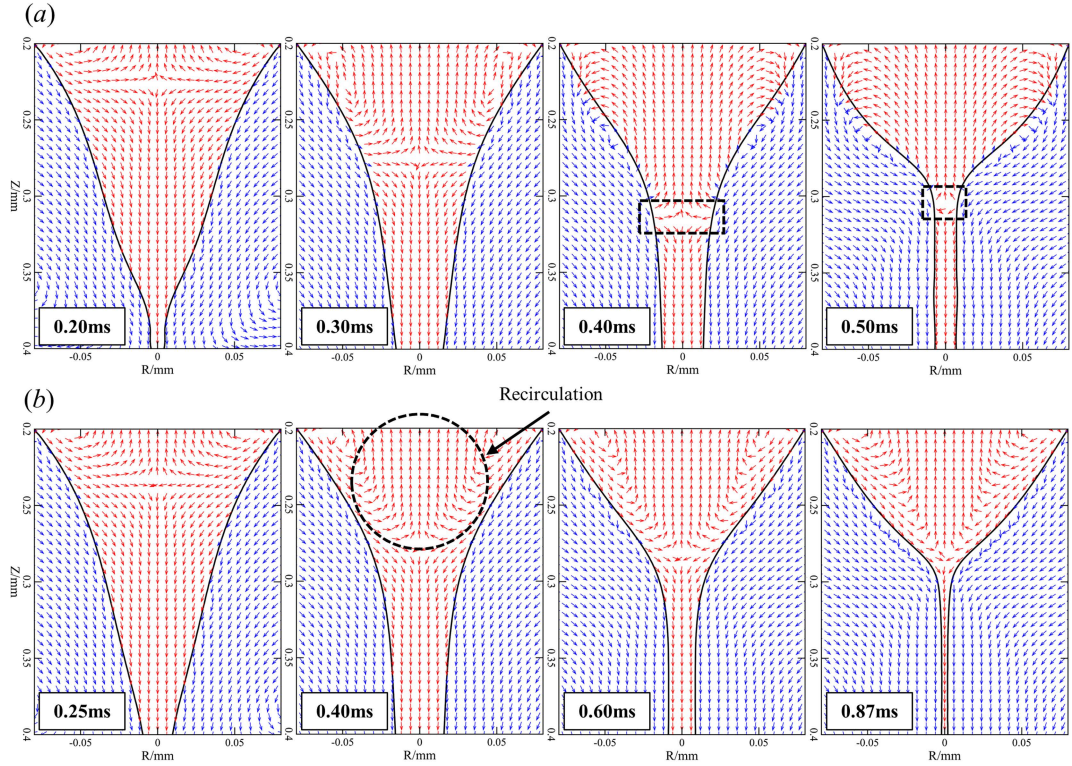

**Fig. 9** The internal flow behaviors of (a) the oscillating cone and (b) the choked jet.

4. Internal flow behaviors

The internal flow behaviors of the oscillating cone and the choked jet exhibit significant differences. In oscillating cone (Fig. 9(a)), the liquid inside the cone and jet initially flowed downward along the axial direction (0.2ms). Subsequently, the flow direction of the liquid inside the cone gradually reversed (0.4ms), while the jet continued to move downward, resulting in flow separation (the black box in Fig. 9(a)). In choked jet (Fig. 9(b)), the liquid inside the cone initially moved downward (0.25ms), and then a recirculation pattern was formed within the Taylor cone (the black circle in Fig. 9(b), 0.4ms). The liquid continuously transported along the surface of the cone to sustain the jet until the instability localized in the cone-jet transition region arose. It can be observed that the internal flow field changes more obviously in the oscillating cone regime, which is attributed to the distinct oscillatory behavior of the Taylor cone between the two regimes.

**B. The dimensionless parameters affecting the ejection regime**

The regime of the pulsating jet can be predicted by the ratio of the minimum flow rate to the input flow rate[19], $Q_m/Q$. When $Q_m/Q > 1$, the regime should be the choked jet; when $Q_m/Q < 1$, the regime should be the oscillating cone. The minimum flow rate of a liquid is solely dependent on

its properties[41,44,45]:

$$Q_m \sim Q_0 \varepsilon_r, \quad \varepsilon_r Re > 1 \tag{8}$$

$$Q_m \sim \frac{Q_0}{Re}, \quad \varepsilon_r Re < 1 \tag{9}$$

Where $Q_0 = \sigma\varepsilon_0/\rho K$ is a characteristic flow rate[46,47], and $Re = (\rho\varepsilon_0\sigma^2/\mu^3 K)^{1/3}$ is the electrohydrodynamic Reynolds number[41]. Therefore, the ratio of the minimum flow rate to the input flow rate can be predicted by:

$$\frac{Q_m}{Q} \sim \frac{Q_0 \varepsilon_r}{Q}, \quad \varepsilon_r Re > 1 \tag{10}$$

$$\frac{Q_m}{Q} \sim \frac{Q_0}{Q Re}, \quad \varepsilon_r Re < 1 \tag{11}$$

We found that $Oh$ (Ohnesorge number) has an important impact on ejection regime and divided the simulation results into three groups for discussion: $Oh = 0.1$, $0.5$, and $1$.

1. $Oh = 0.1$

We first analyze the simulation results of low viscosity liquids ($Oh = 0.1$). Because of $\varepsilon_r Re = (\varepsilon_r^2/\alpha)^{1/3}/Oh$, the inequality $\varepsilon_r Re > 1$ is satisfied easily when the Ohnesorge number is small. Therefore, the dimensionless parameter influencing the ejection regime should be $Q_0\varepsilon_r/Q$. Under the configuration, we have $Q_0\varepsilon_r/Q = \alpha\sqrt{\sigma D^3/\rho}/Q \sim \alpha$, which indicates that the regime is determined by the charge relaxation time. As shown in Fig. 10, only the oscillating cone (OC) and choked jet (CJ) are obtained at this viscosity. The circles and squares correspond to the cases of CJ and OC, respectively. The areas occupied by CJ and OC are denoted by red and grey regions. The boundary between the two regimes is $Q_0\varepsilon_r/Q \approx 10$, indicating that $Q$ is approximately equal to $Q_m$ when $Q_0\varepsilon_r/Q = 10$. The ejection regime is OC and CJ for $Q_0\varepsilon_r/Q < 10 (Q > Q_m)$ and $Q_0\varepsilon_r/Q > 10 (Q < Q_m)$, respectively.

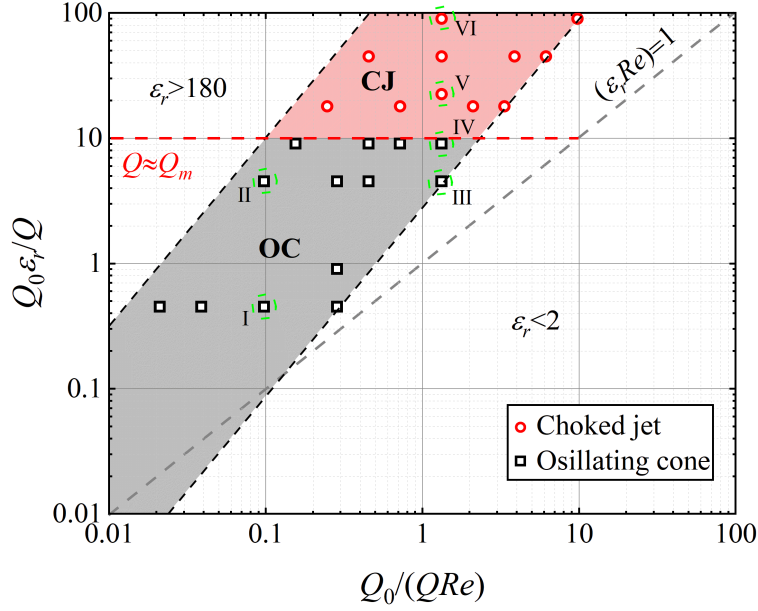

**Fig. 10** The distribution of the ejection regimes of liquids with low viscosity ($Oh=0.1$). The up and down black dashed lines represent $\varepsilon_r=180$ and $\varepsilon_r=2$, respectively. The grey dashed line represents $\varepsilon_r Re=1$. The red dashed line represents the boundary that divides different regimes.

2. $Oh=0.5$

If the liquid viscosity is moderate ($Oh=0.5$), it is necessary to discuss the cases of $\varepsilon_r Re>1$ and $\varepsilon_r Re<1$ respectively. The regimes of all the numerical results of liquids with $Oh=0.5$ are plotted in Fig. 11. The triangles in the figure denote the cases of the stable cone-jet regime (SJ), while the blue region corresponds to the area occupied by SJ. We assume that the boundary between CJ and OC satisfies $Q\approx Q_m$ because a stable jet is impossible to exist when $Q<Q_m$. It can be observed in the figure that $Q_0\varepsilon_r/Q\approx 10$ and $Q_0/(QRe)\approx 1$ divides CJ and SJ at $\varepsilon_r Re>1$ and $\varepsilon_r Re<1$, respectively. Interestingly, the boundary between SJ and OC is $Q_0/(QRe)\approx 0.4$ and irrelevant to the Y-axis. We considered that the transition from SJ to OC is associated with the oscillation of the Taylor cone, which will be discussed further in Section IIIC.

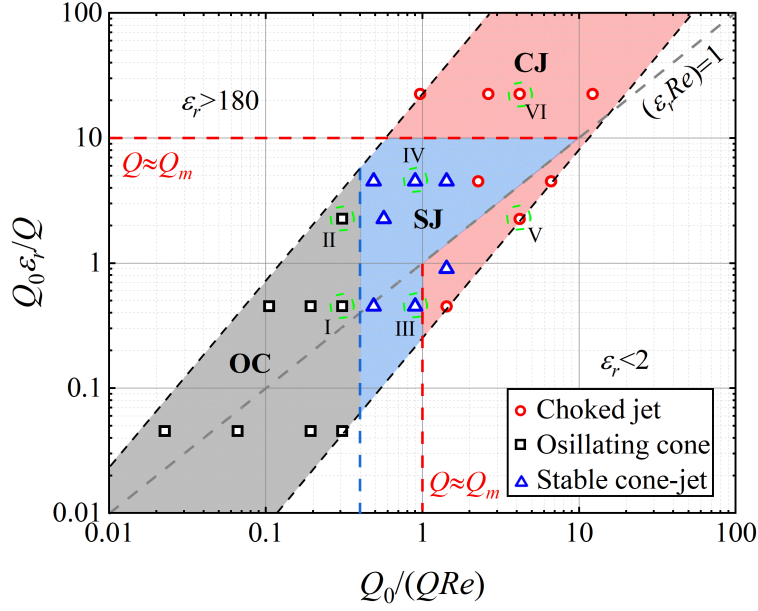

**Fig. 11** The distribution of the ejection regimes of liquids with medium viscosity ($Oh = 0.5$). The up and down black dashed lines represent $\varepsilon_r = 180$ and $\varepsilon_r = 2$, respectively. The grey dashed line represents $\varepsilon_r Re = 1$. The red and blue dashed lines represent the boundaries that divide different regimes.

3. $Oh = 1$

The distribution of the ejection regimes of high viscosity liquids ($Oh = 1$) is shown in Fig. 12. Based on the theoretical prediction, the ratio of the minimal flow rate and the input flow rate is expected to depend on $Q_0 \varepsilon_r / Q$ when $\varepsilon_r Re$ exceeds 1. However, the numerical results indicate that the ejection regimes of high viscosity liquids are only determined by the dimensionless number $Q_0/(QRe)$. Specifically, the ejection regime is classified as CJ, SJ, and OC at $Q_0/(QRe) > 1$, $0.2 < Q_0/(QRe) < 1$, and $Q_0/(QRe) < 0.2$, respectively. The results show that the proposed scaling law about the minimal flow rate of liquids may not be accurate under a high viscosity. Strictly speaking, as the viscosity increases to a certain value, $\varepsilon_r Re = 1$ is no longer a reliable criterion for determining which relationship (Eq. (8) or (9)) should be adopted. We will further discuss this issue in Section IIID.

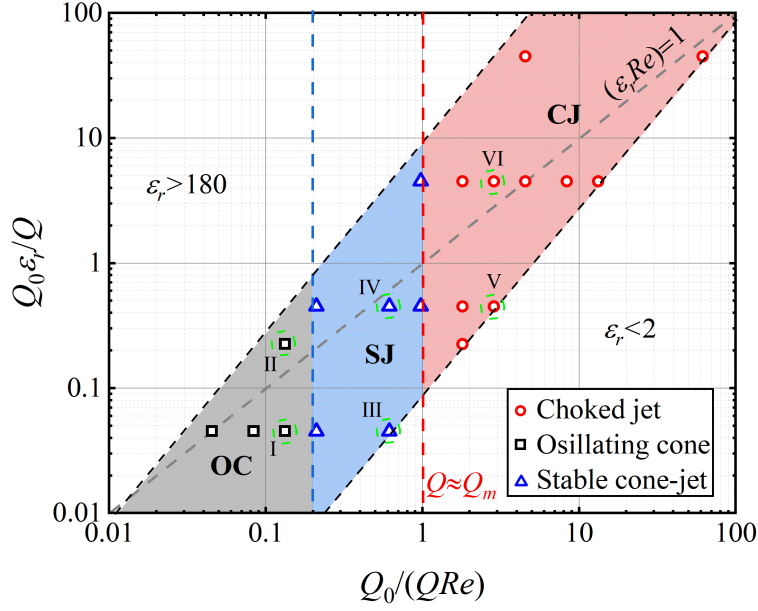

**Fig. 12** The distribution of the ejection regimes of liquids with high viscosity ($Oh=1$). The up and down black dashed lines represent $\varepsilon_r=180$ and $\varepsilon_r=2$, respectively. The grey dashed line represents $\varepsilon_r Re=1$. The red and blue dashed lines represent the boundaries that divide different regimes.

## C. The effects of liquid properties on the cone oscillation in OC

1. Basical theory

The liquid surface can form an oscillating Taylor cone under the action of electric forces, as shown in Fig. 6(b). The previous study[15] has discovered that the oscillation behavior of the cone is similar to the oscillation of a free charged droplet[48,49] and can be simplified to a damped linear oscillator. According to Rayleigh's theory[50], the square of the resonant frequency of the $n$th mode ($\omega_n^2$) of a charged droplet is:

$$\omega_n^2 = \frac{8n(n-1)\sigma}{\rho D^3}\left[(n+2)-\frac{4q^2}{q_R^2}\right] \qquad (12)$$

Where $\rho$, $D$, $q$, and $\sigma$ are the density, diameter, total charge and surface tension of the droplet, respectively; and $q_R=\sqrt{8\pi^2\varepsilon_0\sigma D^3}$ is the Rayleigh limit[50]. The damping ratio of the oscillation is:

$$\zeta = \frac{c}{2m\omega_n} \qquad (13)$$

Where $c$ and $m$ are the damping coefficient and mass, respectively. These results also can be used to predict the oscillating behavior of Taylor cone. If the damping ratio is high, the oscillation is overdamped, thus the amplitude of cone rapidly decays and a stable cone-jet can be formed. Conversely, if the damping ratio is low, the oscillation is underdamped and may cause the interruption of the jetting. In CJ, the mass of cone ($m$ in Eq. (13)) is continuously loss because the input flow rate is lower than the emitted flow rate (equal to the minimal flow rate). Therefore, the damping ratio rises and the cone is expected to be overdamped. In OC, the mass of cone changes slightly and thus the damping ratio at mode $2^{15,51,52}$ can be calculated by selecting $c = 4\pi\mu D(n-1)/n$ and $m = \pi\rho D^3/(4n^2+2n)$:

$$\zeta = \frac{5Oh}{2\sqrt{[1-\left(\frac{q}{q_R}\right)^2]}} \tag{14}$$

The above equation indicates that the damping ratio is related to two dimensionless numbers: the Ohnesorge number $Oh$ and the dimensionless charge $q/q_R$.

2. Effects of viscosity

The change in viscosity corresponds to the change in the Ohnesorge number. As shown in Table. 1, an increase in the Ohnesorge number results in a longer ejection duration, which is caused by the higher damping ratio[53] (according to Eq. (14)). Simultaneously, the difference between the longest and shortest ejection durations at the same viscosity is not significant, indicating a minor influence of conductivity and permittivity on the duration of a single ejection.

**Table. 1** The longest, shortest, and mean ejection duration of OC under different $Oh$ in the simulation results

| $Oh$ | Longest ejection duration /ms | Shortest ejection duration /ms | Mean ejection duration /ms |
|---|---|---|---|
| 0.1 | 0.62 | 0.53 | 0.56 |
| 0.5 | 1.90 | 1.63 | 1.78 |
| 1 | 4.10 | 3.70 | 3.98 |

3. Effects of conductivity

The boundaries between SJ and OC are perpendicular to the X-axis in Fig. 11 and Fig. 12, which means the oscillatory behavior of the cone is purely influenced by $Q_0/(QRe)$ when the viscosity remains constant. Because $Q_0/(QRe) = \sqrt{\sigma D^3/\rho}\, Oh/((\alpha\varepsilon_r)^{2/3}Q) \sim Oh/(\alpha\varepsilon_r)^{2/3}$ under our setting($Q$, $\sigma$, $\rho$ and $D$ are fixed), the variation of $Q_0/(QRe)$ under the same viscosity is caused by the change of $\alpha\varepsilon_r$, which corresponds to the conductivity. According to Eq. (14), we speculate that the conductivity influences the dimensionless charge $q/q_R$ in the cone and further affects the oscillatory behavior.

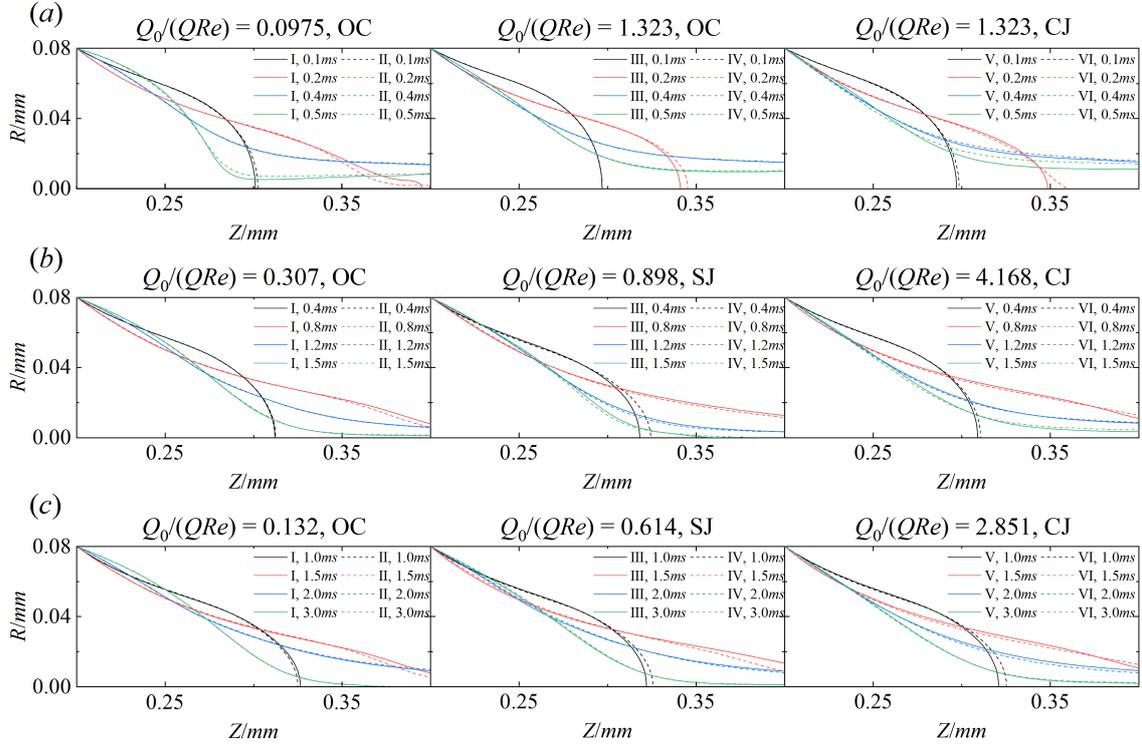

**Fig. 13** The shapes of liquid surfaces of cases with (a) $Oh = 0.1$, (b) $Oh = 0.5$, and (c) $Oh = 1$ at selected moments. For each $Oh$, six cases are selected. The cases with same viscosity, ejection regime, and $Q_0/(QRe)$ are plotted in the same coordinate. All cases are marked in green circles in Fig. 10, Fig. 11, and Fig. 12, and their detailed information is listed in the supplementary material.

To investigate the influence of conductivity on the oscillatory behavior of the meniscus, six cases for each viscosity were selected. The shapes of the liquid surfaces of these cases at different moments are plotted in Fig. 13. For each case, four moments were selected, and the moments chosen for the cases with same $Oh$ were identical. As shown in Fig. 13, the interfaces of the cases in the same coordinates (same $Oh$, regime, and $Q_0/(QRe)$) are coincident at the same moment, indicating a consistency in the oscillation behavior of the cone. Additionally, an increase in $Q_0/(QRe)$ leads to a more noticeable rebound of interface and a more obvious variation in contact angle $\theta$ at all viscosities. For example, in $Oh = 0.1$, the contact angles at 0.1ms, 0.2ms, and 0.5ms in cases I and II ($Q_0/(QRe) = 0.0975$) are 63.5, 54.1, and 73 degrees, respectively; while the contact angles at the same moments in cases III and IV ($Q_0/(QRe) = 1.323$) are 64, 54, and 59.4 degrees, respectively. The different variations in contact angle indicate that a higher conductivity leads to a stronger oscillation of the liquid surface and a greater tendency to form OC.

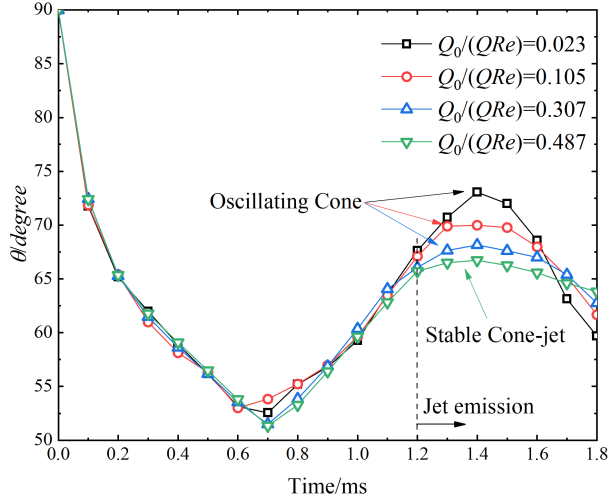

**Fig. 14** The time evolutions of the contact angle $\theta$ for four cases with various $Q_0/(QRe)$ and the same viscosity ($Oh=1$).

The variations of contact angle over time for four high viscosity cases with different $Q_0/(QRe)$ are plotted in Fig. 14 to further analyze the effect of conductivity on the oscillation. Before the jet is emitted, the contact angle rapidly falls to a minimum value and then gradually rises in all cases. At this stage, the variations of contact angle in all cases are consistent, hence their $q/q_R$ should be equal. After the jet generation (1.2ms), the contact angles reach the peak values at about 1.4ms and then decrease. It can be observed that the cone angle of liquids with higher conductivity (lower $Q_0/(QRe)$) have larger variations, which indicates a stronger oscillation. This phenomenon can be explained as follows. The current of the jet can be estimated by the scaling law $I \sim \sqrt{\sigma KQ}$ [41,47,54], thus jets emit from liquids with higher conductivity carry more charges from the Taylor cone to downstream and the amounts of charges ($q/q_R$) inside the cone may be lower. Therefore, the damping ratio is smaller and the oscillation is more obvious for higher conductivity liquids. Moreover, it is worth noting that the contact angles reached positive and negative peak values at the same moments in all cases, implying that changes in conductivity have no significant effect on the oscillation frequency of the cone, as well as the ejection frequency of the oscillating cone regime.

### D. The effects of liquid properties on the dominant resistant forces in CJ

Although the Taylor cone in CJ is stable, the low input flow rate eventually results in the instability in the cone-jet transition region and the interruption of the jetting. In this section, we will discuss the influences of liquid properties on the dominant resistant force[41] that stalls the jet emission in CJ. We adopted the slenderness approximation[46,55,56] and simplified the Navier-stokes equation in the Z-direction to:

$$\rho\left(\frac{dv}{dt}+v\frac{dv}{dz}\right)=F_\sigma+\frac{\varepsilon_0}{2}\frac{d}{dz}[(E_n^o)^2-\varepsilon_r(E_n^i)^2]+\frac{2\varepsilon_0 E_n^o E_z}{R}+\underbrace{\frac{3\mu}{R^2}\frac{d}{dz}\left(R^2\frac{dv}{dz}\right)}_{F_\mu}+\underbrace{\frac{\varepsilon_0}{2}\frac{d}{dz}[(\varepsilon_r-1)E_Z^2]}_{F_p} \quad (15)$$

Where $v$ is the axial velocity, $E_n^o$ is the outer normal component of the electric field at the liquid surface, $E_n^i$ is the inner normal component of the electric field at the liquid surface, $E_z$ is the tangential component of the electric field, and $R$ is the diameter of the liquid column. The first and second terms on the left-hand side of Eq. (15) are referred to as the transient term and inertia, respectively. The right side includes the surface tension, electrostatic force, electric tangential force, viscous force ($F_\mu$), and polarization force ($F_p$), respectively. The viscous force and polarization force are the two possible dominant resistant forces in the cone-jet transition region[41]. According to Gañán-Calvo's theory[41], at $\varepsilon_r Re>1$, the dominant resistant force is the polarization force and $Q_m$ is related to $Q_0\varepsilon_r$; at $\varepsilon_r Re<1$, the viscous force becomes the major resistant force and $Q_m$ is determined by $Q_0/Re$. We only analyze these two forces and discuss whether the dominant resistance in the numerical results corresponds to the theoretical predictions.

The accurate location of the cone-jet transition region can be determined by the current variation along the Z-direction. The total current in the liquid, $I_{total}$, is contributed by the bulk conduction $I_c$ and the surface convection $I_s$:

$$I_{total}=I_c+I_s \quad (16)$$

$$I_c=2\pi\int r\varphi K E_Z dr \quad (17)$$

$$I_s=2\pi\int rv\rho_e dr \quad (18)$$

Where $\varphi$, $\rho_e$, and $r$ are the liquid volume fraction, space charge density, and radial coordinate, respectively. The current distribution in different regions exhibits various characteristics. In the cone region, the current contributed by bulk conduction is higher than the current contributed by surface convection. In the cone-jet transition region, the current of surface convection rises to the same magnitude with bulk conduction, and the quantities of the two current components are comparable ($I_s \approx I_c$). In the jet region, the liquid is fully accelerated, resulting in a high surface convection current.

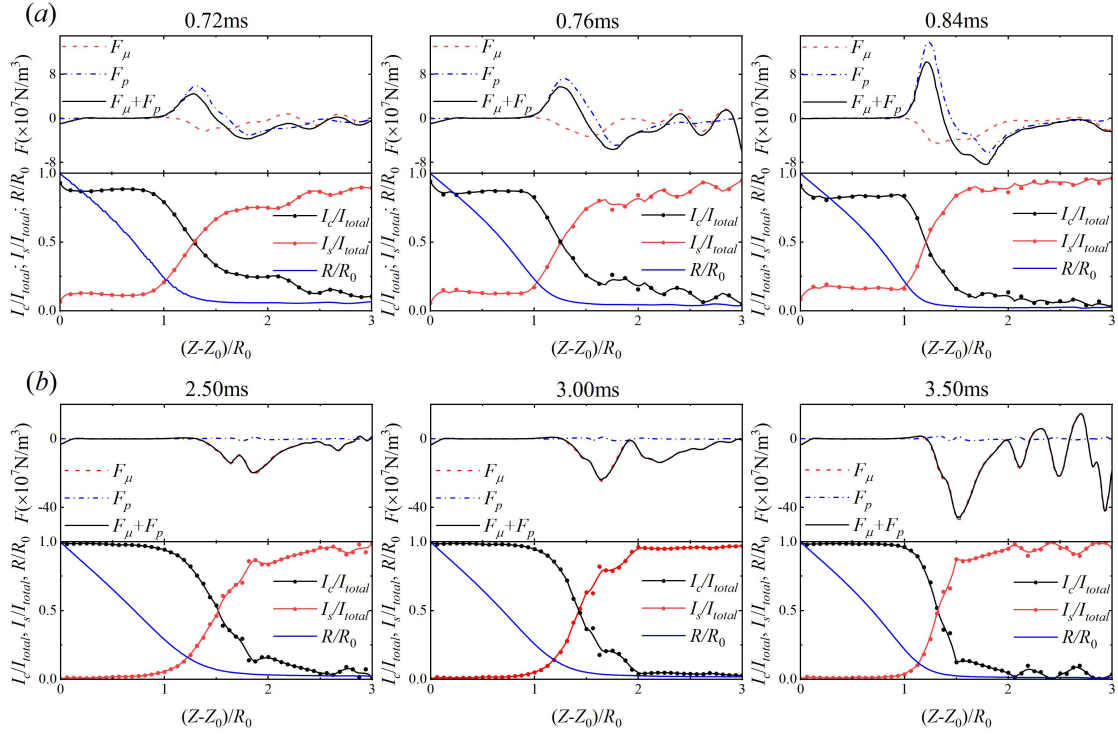

**Fig. 15** The viscous forces, polarization forces, resultant forces of the viscous force and the polarization force, currents, and interface shapes at three different moments of two example cases. (a) $Oh=0.1$, $Q_0\varepsilon_r/Q=45$, and $Q_0/(QRe)=1.3$, the moments from left to right are 0.72ms, 0.76ms, and 0.84ms; (b) $Oh=0.5$, $Q_0\varepsilon_r/Q=2.26$, and $Q_0/(QRe)=4.17$, the moments from left to right are 2.5ms, 3ms, and 3.5ms.

The viscous force, polarization force, current, and interface shape at three moments with $Oh=0.1$, $Q_0\varepsilon_r/Q=45$, and $Q_0/(QRe)=1.3$ are illustrated in Fig. 15(a). The forces are considered positive if the direction is the same as the positive direction of the Z-axis. If the value of a force at a point is greater than zero, the force propels the liquid downstream. Conversely, the force resists the ejection of the jet. It can be observed that the viscous force is always negative throughout the cone-jet transition region and acts as a resistance, while the polarization force only stalls the jet in downstream. The variation trends of the resultant force (sum of the viscous force and polarization force, $F_\mu + F_p$) are identical to the variation trends of the polarization force at three moments.

Additionally, the resultant force ($F_\mu + F_p$) and polarization force reached their negative peak values (indicating maximum hindering role) at approximately the same positions. This suggests that the polarization force predominantly contributes to the resistant effect, particularly downstream of the cone-jet transition region ($1.5 < (Z-Z_0)/R_0 < 2$). In this case ($\varepsilon_r Re > 1$), numerical results are consistent with the theory.

In another case where $Oh = 0.5$, $Q_0\varepsilon_r/Q = 2.26$, and $Q_0/(QRe) = 4.17$ (Fig. 15(b)), the variation curves of the resultant force ($F_\mu + F_p$) and the viscous force are coincident at the three moments, while the variation curves of the polarization force are nearly a straight line. Therefore, the viscous force is the major resistant force in this case, which is the same as the theory. Although we have presented only two cases here, the numerical results of $Oh = 0.1$ and $0.5$ are all consistent with the theoretical predictions of Gañán-Calvo[41].

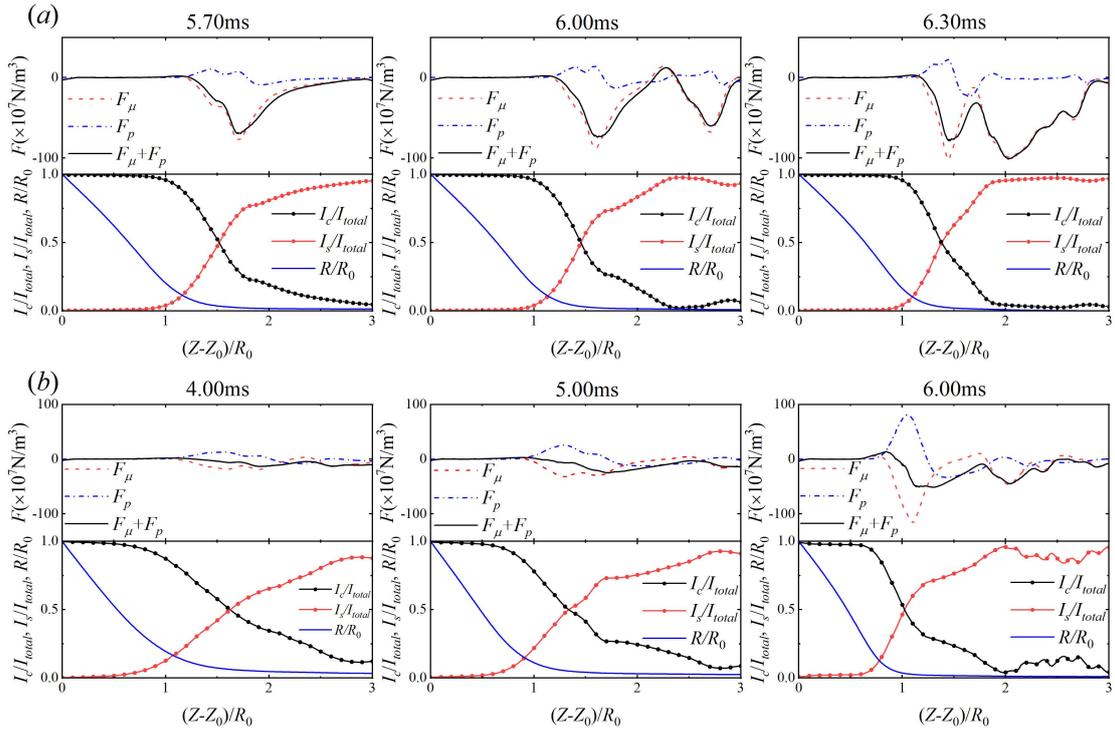

**Fig. 16** The viscous forces, polarization forces, resultant forces of the viscous force and polarization force, currents, and interface shapes at three different moments of two example cases. (a) $Oh = 1$, $Q_0\varepsilon_r/Q = 4.5$, and $Q_0/(QRe) = 2.85$, the moments from left to right are 5.7ms, 6ms, and 6.3ms; (b) $Oh = 1$, $Q_0\varepsilon_r/Q = 45.3$, and $Q_0/(QRe) = 4.53$, the moments from left to right are 4ms, 5ms, and 6ms.

However, the theory cannot exactly predict the dominant resistant force when the viscosity of the liquid exceeds a certain value. We observed that the viscous force always be the dominant resistant force at $Oh=1$. As shown in Fig. 16(a), the variations of the resultant force are nearly identical with the variations of the viscous force at three moments in the case of $Oh=1$, $Q_0\varepsilon_r/Q=4.5$, and $Q_0/(QRe)=2.85$. In another case of $Oh=1$, $Q_0\varepsilon_r/Q=45.3$, and $Q_0/(QRe)=4.53$ (Fig. 16(b)), the polarization force remains positive, while the viscous force remains negative around $I_s=I_c$. Although both forces reach their peak values at the same position ($(Z-Z_0)/R_0=1.1$ at 6ms), the absolute peak values of viscous force are always greater than those of the polarization force, resulting in the negative value of the resultant force at this position. The values of $\varepsilon_r Re$ in the two cases are both larger than 1 (1.58 and 10, respectively). According to the theory, the polarization force should be the dominant resistant force, which conflicts with the numerical results.

In summary, the type of dominant resistant force aligns with the theoretical predictions when $Oh$ = 0.1 and 0.5. However, the viscous force is always the dominant resistant force when $Oh$ = 1, and the boundary between CJ and OC/SJ is only dependent on $Q_0/(QRe)$.

## IV. Conclusion

In this paper, the pulsating phenomenon of the meniscus under an electric field was investigated using numerical simulation. Firstly, a numerical solver based on OpenFOAM was established, and a charge flux restricting step was adopted in the solver to prevent charges from moving into the air phase. The accuracy of the solver was validated by comparing the numerical results and the experiment results from the reference. Subsequently, the ejection processes of liquids with varying properties were simulated and analyzed.

Three distinct ejection regimes were identified in our simulations: oscillating cone, choked jet, and stable cone-jet. The choked jet forms when the input flow rate is lower than the minimal flow rate of the liquid. The oscillatory behavior of the Taylor cone affects the regime if the input flow rate is higher than the minimal flow rate. The oscillating cone appears under strong cone oscillation, while the stable cone-jet forms if the oscillation is quickly stabilized. We discussed the parameters affecting the cone oscillation and the type of dominant resistant force in the cone-jet transition region. Our findings are as follows:

(1) The viscosity and conductivity are the main factors affecting the cone oscillation. A higher viscosity leads to a longer ejection duration and a lower oscillation frequency. A higher conductivity results in a larger amplitude, which increases the possibility of the jet breaking;

(2) The boundary that divides the choked jet and stable cone-jet/oscillating cone regime is $Q_0\varepsilon_r/Q \approx 10$ when the polarization force becomes the dominant resistant force; while it is $Q_0/(QRe) \approx 1$ when the viscous force becomes the main resistance;

(3) The type of the dominant resistant force can be predicted by $\varepsilon_r Re$ if the liquid viscosity is low ($Oh$ = 0.1 and 0.5). However, the dominant resistant force is always the viscous force for liquids with high enough viscosity ($Oh$ = 1).

Our study reveals the influences of liquid properties on the pulsating jet and can be utilized to guide its practical application in EHDP.

## Acknowledgments

This work was supported by the National Natural Science Foundation of China (No. 62103253), the Shanghai Rising-Star Program (No. 21QA1403000), and the Shanghai Science and Technology Committee Natural Science Program (No. 23ZR1423700).

## Author Declarations

### Conflict of interest

The authors have no conflicts to disclose.

## Data Availability

The data that support the findings of this study are available from the corresponding author upon reasonable request.

# Supplementary Material

## I. Numerical solver

The numerical solver used in this study is based on the Taylor-Melcher leaky dielectric model[1,2]. The ions in the liquid are assumed to accumulate at the liquid surface under the action of an electric field. Therefore, the charge density is zero everywhere except the liquid surface. The electric field $E$ can be calculated by the divergence of the electric potential $V$:

$$-\nabla V = E \tag{S1}$$

$$\nabla \cdot (\varepsilon_r E) = \frac{\rho_e}{\varepsilon_0} \tag{S2}$$

where $\varepsilon_0$ and $\varepsilon_r$ are the permittivity of vacuum and the relative permittivity, respectively. The space charge density $\rho_e$ is calculated by the charge conversation equation which is derived from the Nernst-Planck equation[3]:

$$\frac{\partial \rho_e}{\partial t} + \nabla \cdot (u \rho_e) = -\nabla \cdot (KE) \tag{S3}$$

where $u$ and $K$ are the velocity vector and the conductivity, respectively.

We consider the liquid materials are incompressible and immiscible in the air for simplification, and thus the governing equations of the fluid flow are as follows:

$$\nabla \cdot u = 0 \tag{S4}$$

$$\rho \frac{Du}{Dt} = \nabla \cdot [-pI + \mu(\nabla u + (\nabla u)^T)] + \rho g + F_e + F_{st} \tag{S5}$$

where $u$, $p$, $\rho$, $\mu$, and $g$ are the velocity vector, pressure, density, viscosity, and gravitational acceleration, respectively. The electric force $F_e$ is equal to the divergence of the Maxwell stress tensor $T_e$:

$$T_e = \varepsilon_0 \varepsilon_r EE - \frac{1}{2} \varepsilon_0 \varepsilon_r (E \cdot E) I \tag{S6}$$

$$F_e = \nabla \cdot T_e = \rho_e E - \frac{1}{2} E \cdot E \varepsilon_0 \nabla \varepsilon_r \tag{S7}$$

It must be noted that the computation of the electric force is through calculating the divergence of the Maxwell stress tensor rather than using Eq.(S7) directly. The research of López-Herrera et al.[4] shows that the results are much more accurate when computed using the divergence form.

The volume of fluid (VOF) method is used to track the interface of the liquid. In the VOF method, a volume fraction of liquid $\varphi$ is adopted to describe the distribution of different fluids:

$$\begin{cases} \varphi = 1, & liquid \\ 0 < \varphi < 1, & interface \\ \varphi = 0, & air \end{cases} \tag{S8}$$

The MULES-VOF (Multi-dimensional Universal Limiter for Explicit Solution) approach is adopted in the built-in solver interFOAM of OpenFOAM software, which solves the following equation:

$$\frac{\partial \varphi}{\partial t} + \nabla \cdot (\boldsymbol{u}\varphi) + \nabla \cdot (c\boldsymbol{u}_{norm}\varphi(1-\varphi)) = 0 \tag{S9}$$

where $\boldsymbol{u}_{norm}$ is the norm velocity at the interface. The role of the third term in the equation is to compress the thickness of the interface layer in order to simulate real situations as far as possible. The coefficient $c$ is used to control the effect of interfacial compression and is normally set to 1.

The physical properties of the fluid are varying in the computational domain according to the value of $\varphi$:

$$\rho = \varphi \rho_l + (1-\varphi) \rho_a \tag{S10}$$

$$\mu = \varphi \mu_l + (1-\varphi) \mu_a \tag{S11}$$

$$\varepsilon_r = \varphi \varepsilon_l + (1-\varphi) \varepsilon_a \tag{S12}$$

$$K = \varphi K_l + (1-\varphi) K_a \tag{S13}$$

The subscripts $l$ and $a$ represent the corresponding properties of liquid and air, respectively.

The surface tension force $\boldsymbol{F}_{st}$ is computed using the Continuum-Surface-Force (CSF) approach:

$$\boldsymbol{F}_{st} = -\sigma \left[ \nabla \cdot \left( \frac{\nabla \varphi}{|\nabla \varphi|} \right) \right] \nabla \varphi \tag{S14}$$

Where $\sigma$ is the surface tension coefficient of the liquid. The flow chart of the solver in this research is demonstrated in Fig.S1.

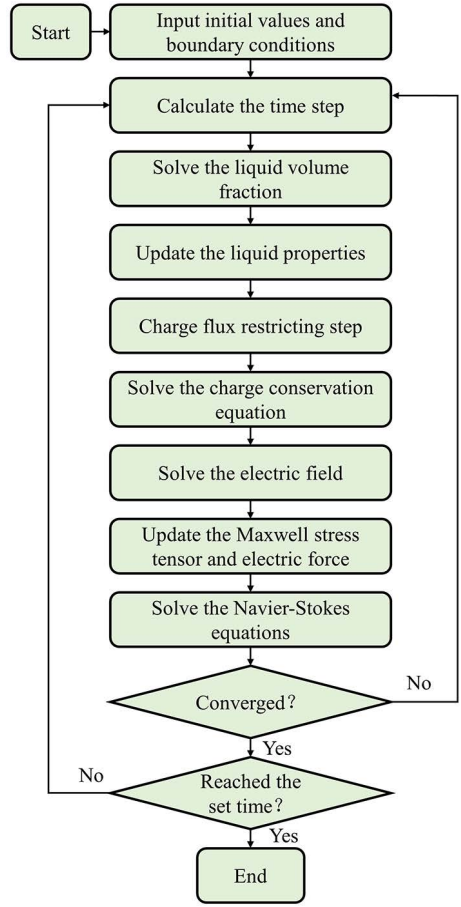

**Fig.S1** The flow chart of the solver.

## II. Mesh independent research

We calculated the amounts of total charges in the drops at 0.1ms under different minimum mesh sizes:

$$Q_{total} = \int_{\Omega} 2\pi r \rho_e \, d\Omega \tag{S15}$$

Where $r$ is the radial coordinate, $\rho_e$ is the charge density at a certain point in the space, and $\Omega$ represents the computational domain. As shown in Fig.S2, the amounts of total charges remain almost unchanged when the minimum mesh size is below 1.6μm. Therefore, the numerical solutions are independent when using a mesh that minimum size is small than 1.6μm. In our study, we constructed a mesh whose minimum size is 1μm, as depicted in Fig.S3.

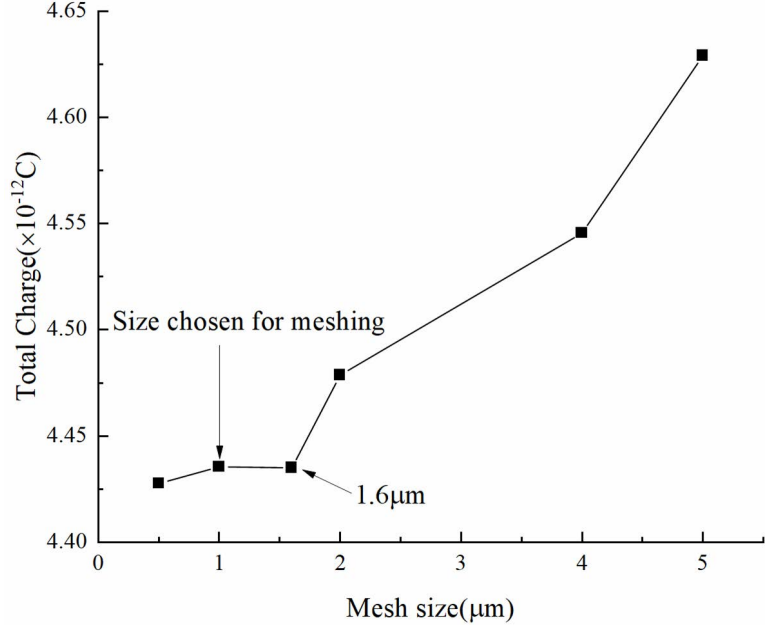

**Fig.S2** The amount of total charges in the liquid drop at 0.1ms of the different mesh sizes.

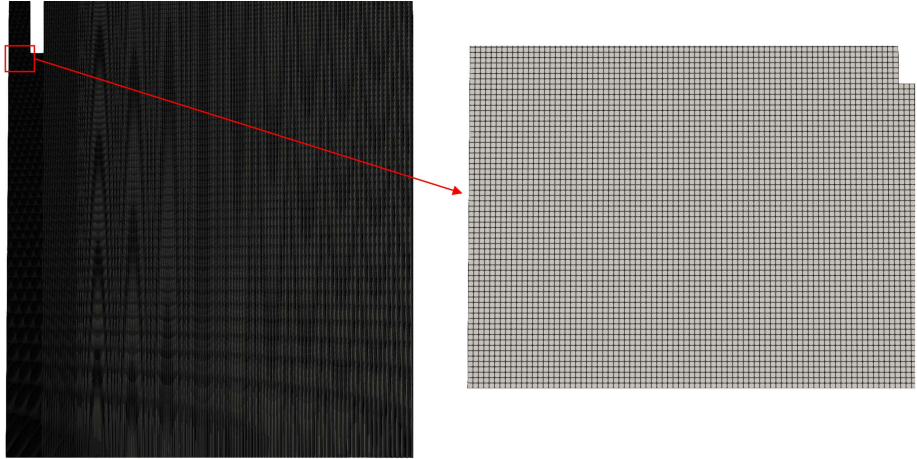

**Fig.S3** The mesh used for simulations. The minimal mesh size is 1μm.

## III. The list of the parameters of liquids

**Table.S1** The list of the dimensionless parameters of liquids at $Oh = 0.1$.

|   | $\alpha$ | $\varepsilon_r$ | $Q_0/(ReQ)$ | $Q_0\varepsilon_r/Q$ | $\varepsilon_r Re$ |
|---|---|---|---|---|---|
| 1 | 1000 | 2 | 0.28509 | 0.45255 | 1.5874 |
| 2 (I) | 1000 | 10 | 0.097499 | 0.45255 | 4.6416 |
| 3 | 1000 | 40 | 0.038692 | 0.45255 | 11.696 |
| 4 | 1000 | 100 | 0.021005 | 0.45255 | 21.544 |
| 5 | 500 | 4 | 0.28509 | 0.9051 | 3.1748 |
| 6 (III) | 100 | 2 | 1.3233 | 4.5255 | 3.42 |
| 7 | 100 | 10 | 0.45255 | 4.5255 | 10 |
| 8 | 100 | 20 | 0.28509 | 4.5255 | 15.874 |
| 9 (II) | 100 | 100 | 0.097499 | 4.5255 | 46.416 |
| 10 (IV) | 50 | 4 | 1.3233 | 9.051 | 6.8399 |
| 11 | 50 | 10 | 0.71838 | 9.051 | 12.599 |
| 12 | 50 | 20 | 0.45255 | 9.051 | 20 |
| 13 | 50 | 100 | 0.15477 | 9.051 | 58.48 |
| 14 | 25 | 2 | 3.3344 | 18.102 | 5.4288 |
| 15 | 25 | 4 | 2.1005 | 18.102 | 8.6177 |
| 16 | 25 | 20 | 0.71838 | 18.102 | 25.198 |
| 17 | 25 | 100 | 0.24568 | 18.102 | 73.681 |
| 18 (V) | 20 | 10 | 1.3233 | 22.627 | 17.1 |
| 19 | 10 | 2 | 6.142 | 45.255 | 7.3681 |
| 20 | 10 | 4 | 3.8692 | 45.255 | 11.696 |
| 21 | 10 | 20 | 1.3233 | 45.255 | 34.2 |
| 22 | 10 | 100 | 0.45255 | 45.255 | 100 |
| 23 | 5 | 2 | 9.7499 | 90.51 | 9.2832 |
| 24 (VI) | 5 | 40 | 1.3233 | 90.51 | 68.399 |

**Table.S2** The list of the dimensionless parameters of liquids at $Oh = 0.5$.

| | $\alpha$ | $\varepsilon_r$ | $Q_0/(ReQ)$ | $Q_0\varepsilon_r/Q$ | $\varepsilon_r Re$ |
|---|---|---|---|---|---|
| 1 | 10000 | 2 | 0.3071 | 0.045255 | 0.14736 |
| 2 | 10000 | 4 | 0.19346 | 0.045255 | 0.23392 |
| 3 | 10000 | 20 | 0.066163 | 0.045255 | 0.68399 |
| 4 | 10000 | 100 | 0.022627 | 0.045255 | 2 |
| 5 | 1000 | 2 | 1.4254 | 0.45255 | 0.31748 |
| 6 (III) | 1000 | 4 | 0.89797 | 0.45255 | 0.50397 |

| | | $\varepsilon_r$ | $Q_0/(ReQ)$ | $Q_0\varepsilon_r/Q$ | $\varepsilon_r Re$ |
|---|---|---|---|---|---|
| 7 | 1000 | 10 | 0.48749 | 0.45255 | 0.92832 |
| 8 (I) | 1000 | 20 | 0.3071 | 0.45255 | 1.4736 |
| 9 | 1000 | 40 | 0.19346 | 0.45255 | 2.3392 |
| 10 | 1000 | 100 | 0.10503 | 0.45255 | 4.3089 |
| 11 | 500 | 4 | 1.4254 | 0.9051 | 0.63496 |
| 12 (V) | 200 | 2 | 4.168 | 2.2627 | 0.54288 |
| 13 | 200 | 40 | 0.56569 | 2.2627 | 4 |
| 14 (II) | 200 | 100 | 0.3071 | 2.2627 | 7.3681 |
| 15 | 100 | 2 | 6.6163 | 4.5255 | 0.68399 |
| 16 | 100 | 10 | 2.2627 | 4.5255 | 2 |
| 17 | 100 | 20 | 1.4254 | 4.5255 | 3.1748 |
| 18 (IV) | 100 | 40 | 0.89797 | 4.5255 | 5.0397 |
| 19 | 100 | 100 | 0.48749 | 4.5255 | 9.2832 |
| 20 | 20 | 4 | 12.187 | 22.627 | 1.8566 |
| 21 (VI) | 20 | 20 | 4.168 | 22.627 | 5.4288 |
| 22 | 20 | 40 | 2.6257 | 22.627 | 8.6177 |
| 23 | 20 | 180 | 0.96331 | 22.627 | 23.489 |

**Table.S3** The list of the dimensionless parameters of liquids at $Oh=1$.

| | $\alpha$ | $\varepsilon_r$ | $Q_0/(ReQ)$ | $Q_0\varepsilon_r/Q$ | $\varepsilon_r Re$ |
|---|---|---|---|---|---|
| 1 (III) | 10000 | 2 | 0.6142 | 0.045255 | 0.073681 |
| 2 | 10000 | 10 | 0.21005 | 0.045255 | 0.21544 |
| 3 (I) | 10000 | 20 | 0.13233 | 0.045255 | 0.342 |
| 4 | 10000 | 40 | 0.08336 | 0.045255 | 0.54288 |
| 5 | 10000 | 100 | 0.045255 | 0.045255 | 1 |
| 6 | 2000 | 2 | 1.7959 | 0.22627 | 0.12599 |
| 7 (II) | 2000 | 100 | 0.13233 | 0.22627 | 1.71 |
| 8 (V) | 1000 | 2 | 2.8509 | 0.45255 | 0.15874 |
| 9 | 1000 | 4 | 1.7959 | 0.45255 | 0.25198 |
| 10 | 1000 | 10 | 0.97499 | 0.45255 | 0.46416 |
| 11 (IV) | 1000 | 20 | 0.6142 | 0.45255 | 0.73681 |
| 12 | 1000 | 100 | 0.21005 | 0.45255 | 2.1544 |
| 13 | 100 | 2 | 13.233 | 4.5255 | 0.342 |
| 14 | 100 | 4 | 8.336 | 4.5255 | 0.54288 |
| 15 | 100 | 10 | 4.5255 | 4.5255 | 1 |
| 16 (VI) | 100 | 20 | 2.8509 | 4.5255 | 1.5874 |
| 17 | 100 | 40 | 1.7959 | 4.5255 | 2.5198 |
| 18 | 100 | 100 | 0.97499 | 4.5255 | 4.6416 |
| 19 | 10 | 2 | 61.42 | 45.255 | 0.73681 |
| 20 | 10 | 100 | 4.5255 | 45.255 | 10 |